\documentclass[8pt,eqsecnum,preprint2]{aastex}
\bibpunct{[}{]}{,}{a}{}{;}
\newcommand{\hi}{H{\scriptsize{I}}}
\newcommand{\br}[1]{\left(#1\right)}
\newcommand{\kms}{\,\mathrm{km}/\mathrm{s}}

\newcommand{\kpc}{\,\mathrm{kpc}}
\newcommand{\mpc}{\,\mathrm{Mpc}}
\newcommand{\ud}[1]{\mathrm{d}#1}
\renewcommand{\sq}[1]{\left[#1\right]}
\newcommand{\vp}{\mathcal{P}}
\renewcommand{\exp}[1]{e^{#1}}
\renewcommand{\clearpage}{{}}
\shorttitle{Iterative spectral method for spiral galaxies}
\shortauthors{J. Ja{\l}ocha, {\L}. Bratek, M. Kutschera}
\begin{document}
\title{Is dark matter present in NGC 4736?\\
An iterative spectral method for finding mass distribution\\
in spiral galaxies}
\author{Joanna Ja{\l}ocha, {\L}ukasz Bratek, Marek Kutschera\altaffilmark{1}}
\affil{The Henryk Niewodnicza{\'n}ski Institute of Nuclear
Physics, Polish Academy of Science,\\ Radzikowskego 152, 31-342
Krak{\'o}w, Poland} \altaffiltext{1}{Jagellonian University,
Institute of Physics, Reymonta 4, 30-059 Krak{\'o}w, Poland.}
\begin{abstract}
An iterative method for reconstructing mass distribution in spiral
galaxies using a thin disk approximation is developed. As an
example, the method is applied to galaxy NGC 4736; its rotation
curve does not allow one to employ a model with a massive
spherical halo. We find a global mass distribution in this galaxy
(without non-baryonic dark matter) that agrees perfectly with the
high resolution rotation curve of the galaxy. This mass
distribution is consistent with the $I$-band luminosity profile
with the mean mass-to-light ratio $M/L_I=1.2$, and also agrees
with the amount of hydrogen observed in the outermost regions of
the galaxy. We predict the total mass of the galaxy to be only
$3.43\times10^{10}M_{\sun}$. It is very close to the value
predicted by the modified gravity models and much less than the
currently accepted value of $5.0\times10^{10}M_{\sun}$ (with
$\approx70\%$ of the mass in a dark matter halo).
\end{abstract}


\section{Introduction}
Determination of dynamical masses of spiral galaxies is a decisive
step in establishing empirically the clustering scale of dark
matter. The Cold Dark Matter model predicts a hierarchy of scales,
from a dwarf sub-halo scale of $10^6$ of solar masses up to a
galaxy cluster size. There is a strong evidence from the $X$-ray
observations of galaxy clusters, and from detecting dark matter
with gravitational lensing \citep{bib:20} for a $1\mpc$ scale,
which is the galaxy cluster scale. At the galactic size scale (of
the order of $10-50\kpc$), there are contradictory results. Dark
matter was not observed in some elliptical galaxies
\citep{bib:21}. As for the spiral galaxies, it was pointed out
that the amount of dark matter in the disk of the Milky Way galaxy
near the Sun is insignificant \citep{bib:30}. Moreover, mass
estimates of spiral galaxies are model-dependent \citep{bib:22}.

In the following we propose an Iterative Spectral Method to
reconstruct mass distribution in the spiral galaxies. The method
is based on the thin disk model \citep{bib:toomre,bib:nord} which
offers the possibility to obtain mass density distribution from
the observed rotation curve \citep{bib:23}. With the Iterative
Spectral Method this reconstruction is unique, provided there is
observational data about the hydrogen density in the regions
outside of rotation velocity measurements. Without such data one
can also extrapolate hydrogen measurements by assuming its falloff
profile that would be physically acceptable. One should also take
into account the abundance of helium which is proportional to that
of hydrogen. In this way we obtain a complete set of data in the
complementary regions of a given galaxy. Complete data are
required to eliminate cutoff errors that normally result from
cutting-off integration in an integral that relates rotational
velocity to disk surface mass density at the radius where
measurements of a rotation curve end. We shall call the radius the
'cutoff radius' and denote it by $R$.

In fact, in early days of galactic research the thin disk model
was applied to infer the galactic masses from the measured
rotation curves. However, large errors of measurements of
rotational velocities practically made the surface mass density
reconstruction impossible \citep{bib:23}. To avoid uncertainties
in the thin disk mass reconstruction it became customary to a
priori divide spiral galaxies into three major components: bulge,
disk and spherical halo. Then a several parameter fit was made
determining the properties of those components. Although this
approach is physically appealing, the amount of arbitrariness in
such a division is unacceptable. In particular, the most massive
component of this construction - the spherical halo
 - stands on the weakest empirical footing.

In this paper we address the problem of determination of masses of
spiral galaxies with rotation curves violating the sphericity
condition at large radii. One expects that the disk model should
approximate mass distribution in such galaxies better than models
assuming a massive spherical halo. The reconstructed surface mass
density found for galaxy NGC 4736 approaches the hydrogen density
at the outskirts of the galaxy, the rotation curve is perfectly
matched in the measured range, and its continuation is predicted
further out into the Keplerian regime. Thus the disk model
performs much better in this case than models including the
spherical dark halo.

\begin{figure}[h!!!!]\centering
\includegraphics[width=\columnwidth]{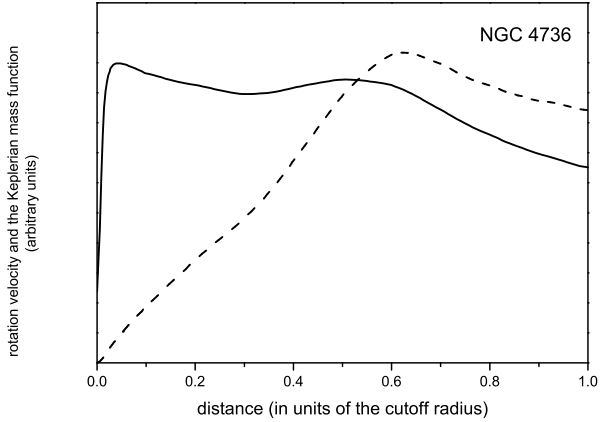}
\caption{\label{fig:f1} Negative sphericity test for spiral galaxy
NGC 4736. The figure shows experimental rotation curve (solid
line) and the corresponding Keplerian mass function defined by
$M_{K}(\rho)=G^{-1}\rho\,v^2(\rho)$ (dashed line). The Keplerian
mass function is not increasing at all distances, therefore a
massive spherical halo in the approximation of circular orbits is
excluded and should be replaced with an extended
flattened component. }\end{figure}

To see why the assumption of the existence of spherical halo seems
to be unjustified for a class of spiral galaxies, we consider as
an example the galaxy NGC 4736. It was found (\citet{bib:kent})
that most of the mass ($68\%$) in this galaxy is concentrated in a
spherical dark halo, mainly at larger radii. At these distances
surfaces of constant gravitational potential are almost spherical.
Therefore one would expect in this region the rotation curve
predicted by the Kent's analysis to satisfy the sphericity
condition. For a spherically symmetric mass distribution the
Keplerian mass function, $M_K(\rho)\equiv\,G^{-1}\,v^2(\rho)\rho$,
corresponding to rotational velocity $v(\rho)$, should be a
nondecreasing function of the radial distance from the center
$\rho$, that is,
$$v^2(\rho_1)\rho_1\leq{}v^2(\rho_2)\rho_2,\qquad
\mathrm{if \ only} \qquad \rho_1\leq\rho_2.$$ The experimental
rotation curve of NGC 4736 violates this condition for
$\rho\gtrsim6\kpc$, see figure \ref{fig:f1}, whereas the
best-fitting solution for the rotation curve found by Kent
satisfies this condition. At large radii the best-fitting
rotational velocity  is also greater from the observed  by a
factor of $\approx1.2$. The total mass should be therefore
$\approx5.9\times10^{10}M_{\sun}/(1.2)^2=4.1\times10^{10}M_{\sun}$
rather than $5.9\times10^{10}M_{\sun}$ as found in
\citep{bib:kent} (at the currently accepted distance $5.1\mpc$ to
NGC 4736 the mass is even less,
$\approx3.5\times10^{10}M_{\sun}$).
%
%
%
This shows that the mass distribution in the galaxy cannot be
dominated by a spherical component and that it would be more
appropriate to assume more extended disk components in similar
galaxies in place of massive spherical halos.

The global disk model has been used recently \citep{bib:5},
however the procedure of cutting-off integrals applied therein
lead to overestimation of surface density at larger radii for
galaxies with declining rotation curves. For example, surface
density in the galaxy NGC 4736 could be reliably determined only
within a radius of less than $6\kpc$. But the experimental data
available for this galaxy enable one to find a much better
determination of mass distribution as far as $16\kpc$ from the
center. An improved approach was presented in \citep{bib:5a},
where rotation curves were assumed Keplerian in the unobserved
regions. This procedure is still unsatisfactory, as velocities of
test particles on circular orbits encircling a disk-like object
need not be Keplerian. Besides no comparison with the observed
mass density has been done.
\section{Iterative Spectral Method}

Let $v_R=v(R)$ denote the last measured value of rotational
velocity at cutoff radius $R$. We define also a dimensionless
function $u(x)=v(xR)/v_R$ for describing rotation curve, where
$x=\rho/R$ measures distances in units of the cutoff radius (we
work in cylindrical coordinates $\br{\rho,\phi,z}$).

Below we shall derive equations that are fundamental for the
iteration method we present later on.
\subsection{Basic equations of disk model}
Let the $z=0$ plane represent the galactic disc. It is assumed
that most of galactic mass is concentrated in its vicinity. A
$z$-symmetric and cylindrically symmetric vacuum gravitational
potential is given by
$$
\Phi(\rho,z)=-2\pi{}v_R^2\int\limits_0^{\infty}\hat{\sigma}(\omega)
J_0\br{\omega\frac{\rho}{R}} \exp{-\omega\frac{|z|}{R}}
\ud{\omega}. $$
A discontinuity of $\frac{1}{2\pi{}G}\vec{\nabla}\Phi$ in the
direction normal to the $z=0$ plane  is interpreted as a surface
mass density $\sigma(\rho)$ of the galactic disk
\begin{equation}\label{eq:sig_from_hat_sig}
\sigma(\rho)=\frac{v_R^2}{GR
}\int\limits_0^{\infty}\omega\hat{\sigma}(\omega)
J_0\br{\omega\frac{\rho}{R}}\ud{\omega},\end{equation} thus
$\hat{\sigma}(\omega)$ is the Fourier-Bessel transform of
$\sigma(\rho)$
\begin{equation}
\label{eq:sigmaintegral}
\hat{\sigma}(\omega)=\frac{GR}{v_R^2}\int\limits_0^{\infty}
x\,\sigma(R\,x)J_0(\omega\,{x})\ud{x}.
\end{equation}
Modelling of galactic rotation curves usually assumes that orbits
of stars are circular and concentric, then
$\left.v^2(\rho)\right|_{z=0}=\rho\,\partial_{\rho}\Phi(\rho,0)$,
therefore,
\begin{equation}\label{eq:u2}
u^2(x)=2\pi\,{x}\int\limits_0^{\infty}\omega\,
\hat{\sigma}(\omega)J_1(\omega{x})\ud{\omega}.
\end{equation}
By substituting (\ref{eq:sigmaintegral}) to (\ref{eq:u2}) one can
show that\footnote{We used the following identities
\citep{bib:15}: $\omega^{-1}\partial_x\br{xJ_1(\omega{x})}
=xJ_0(\omega{x})$ and
\begin{eqnarray*}&&\int_0^{\infty}\ud{\omega}J_1(\omega{\xi})
J_1(\omega{x})=\\&&=\frac{2}{\pi}\left\{
\begin{array}{lll}
\sq{K\br{\xi/x}-E\br{\xi/x}}\xi^{-1},& 0<\xi<x,\\
\sq{K\br{x/\xi}-E\br{x/\xi}}x^{-1},& 0<x<\xi,
\end{array}\right.\end{eqnarray*} where
the complete elliptic functions are defined by
$$
K(\kappa)=\!\!\!\int\limits_{0}^{\pi/2}\!\!\frac{\ud{\phi}}{
\sqrt{1-\kappa^2\sin^2{\phi}}}\ ,\quad
E(\kappa)=\!\!\!\int\limits_{0}^{\pi/2}\!\!\ud{\phi}
\sqrt{1-\kappa^2\sin^2{\phi}}\ .$$}
\begin{eqnarray}v^2(\rho)=4\,G\,\rho\,\cdot\,\vp\left(
\int\limits_{0}^{\rho}\sigma(\chi)\frac{\chi{}E
\br{\frac{\chi}{\rho}}}{\rho^2-\chi^2}\,\ud{\chi}\right.\nonumber\\
\left.- \int\limits_{\rho}^{\infty}\sigma(\chi)\sq{
\frac{\chi^2E\br{\frac{\rho}{\chi}}}{\rho\br{\chi^2-\rho^2}} -
\frac{K\br{\frac{\rho}{\chi}}}{\rho}}\ud{\chi}\right)
,\label{eq:vfromsig}\end{eqnarray}
where $\vp$ stands for the "principal value integral". The simple
pole $\chi=\rho$ dominates the logarithmic divergence in
$K(\rho/\chi)$, thus the principal value integral is easily
tractable numerically and can be calculated with high accuracy.
The inverse of the above integral reads\footnote{Equation
(\ref{eq:sigmamoja}) is deliberately presented in the form
containing only squares of rotational velocity. Much more familiar
is its equivalent form given in the \cite{bib:23} handbook, and
therefore commonly used, however it has the deficiency of
containing derivatives of $v^2(\rho)$ which are determined
experimentally with very low accuracy.}
\begin{eqnarray}\label{eq:sigmamoja}\sigma(\rho)=\frac{1}{\pi^2G}
\vp\left(\int\limits_\rho^{\infty}v^2(\chi)
\frac{E\br{\frac{\rho}{\chi}}
}{\chi^2-\rho^2}\,\ud{\chi}\right.\nonumber\\
\left. + \int\limits_0^\rho
v^2(\chi)\sq{\frac{K\br{\frac{\chi}{\rho}}}{ \rho\
\chi}-\frac{\rho}{\chi}
\frac{E\br{\frac{\chi}{\rho}}}{\rho^2-\chi^2}}\ud{\chi}\right),\end{eqnarray}
Unfortunately, this nice formula is not very useful for
determining $\sigma(\rho)$ directly from $v(\rho)$, as the
observed rotation curves have naturally a cut-off, and therefore
the integration cannot be carried out.\footnote{ however, the
relation serves as a consistency check of our final results we
obtain iteratively later on} For example, the closer $\rho$ is to
cutoff radius $R$, the more $\sigma(\rho)$ is underestimated by
cutting off the integration at $R$. In order to overcome this
unpleasant feature of disk geometry one may use another,
complementary set of the experimental data concerning mass
distribution at $\rho>R$. A global mass density and the
corresponding global rotation curve can be found iteratively that
would both satisfy the relations (\ref{eq:vfromsig}) and
(\ref{eq:sigmamoja}) and be consistent with the observations.

\subsection{A series approximation of the surface density}\label{sec:spect_decomp}

Below we shall find a first approximation to $\sigma(\rho)$
directly from the observed part of rotation curve.

A function $u^2(x)/x$ defined on the unit interval $x\in(0,1)$,
can be expressed in terms of a series of independent functions. In
particular, if $u^2(x)/\sqrt{x}$ is integrable over the interval,
a complete set of orthogonal functions on $x\in(0,1)$ can be
constructed from the Bessel function $J_1$ \citep{bib:lebiediev}.
Then
\begin{eqnarray}
&\frac{u^2(x)}{x}=\sum\limits_{k}\hat{\sigma}_k J_1\br{\omega_kx},&\nonumber\\
\label{eq:vseries} &0<x<1,\quad J_0(\omega_k)=0,\quad\omega_k>0.&
\end{eqnarray}
The summation is taken over all consecutive zeros $\omega_k>0$ of
the Bessel function $J_0$. By multiplying both sides by
$x\,J_1(\omega_m\,x)$ and integrating over $x\in(0,1)$, we find
that
\begin{equation}
\label{eq:expanscoeff}
\hat{\sigma}_k=\frac{\omega_k}{J_1(\omega_k)}\mu_k,\quad
\mu_k=\frac{2}{\omega_k}\int\limits_0^1u^2(x)
\frac{J_1(\omega_kx)}{J_1(\omega_k)}\ud{x}.\end{equation} The
amount of information encoded in the enumerable sequence
$\hat{\sigma}_{k}$ is insufficient for determining the transform
$\hat{\sigma}(\omega)$ which is a continuum of numbers. Therefore,
$\sigma(\rho)$ cannot be determined accurately from the observed
rotation curve alone, even in the region $\rho<R$. However, the
elements of the sequence can be still linearly combined to
approximate $\hat{\sigma}(\omega)$. Note, that every such
approximation is equivalent to choosing a particular extrapolation
of a rotation curve beyond its cutoff radius.

There are some mathematical subtleties regarding this
extrapolation. Let us assume first that we put $v(\rho)=0$ for
$\rho>R$ and $v(\rho)=v_R/2$ at a single point $\rho=R$. Then we
can use the representation (\ref{eq:vseries}) for $x<1$, and for
$x>1$ we set $u=0$. From the inverse transform of (\ref{eq:u2}),
which reads
\begin{equation}
\label{eq:sigmaomega} \hat{\sigma}(\omega)=
\frac{1}{2\pi}\int\limits_0^{\infty}u^2(x)J_1(\omega{x})\ud{x},
\end{equation}we then obtain
a continuous spectrum\footnote{Since
$J_0(\omega)\approx{}J_1(\omega_k)(\omega_k-\omega)$ when
$\omega\approx\omega_k$, the components under the summation sign
are continuous functions of $\omega$}
\begin{equation}\label{eq:contin}\hat{\sigma}(\omega)=\frac{1}{2\pi}\sum\limits_k\hat{\sigma}_k\frac{\omega
{}J_0(\omega)J_1(\omega_k)}{\omega_k^2-\omega^2}.\end{equation}
However, for handling the observational data more tractable are
discrete spectra. For instance, by extending formally
(\ref{eq:vseries}) to the region $x>1$, we get
\begin{equation}\label{eq:discretespect}
\hat{\sigma}(\omega)=\frac{1}{2\pi}\sum\limits_k\hat{\sigma}_k\frac{
\delta(\omega-\omega_k)}{\omega}.\end{equation} Now, the
corresponding $u^2(x)$, calculated from equation (\ref{eq:u2}),
 can behave quite odd for $x>1$ and become even negative!
Nevertheless, both (\ref{eq:contin}) and (\ref{eq:discretespect}),
when substituted to (\ref{eq:u2}), lead to the same $u^2(x)$ for
any $x\in(0,1)$. Although the corresponding surface mass densities
differ from each other also for $\rho<R$, the difference is, in
practice, negligible close to the centre. We may therefore accept
(\ref{eq:discretespect}) as a first approximation of
$\hat{\sigma}(\omega)$. By substituting (\ref{eq:discretespect})
to (\ref{eq:sig_from_hat_sig}) we obtain
\begin{equation}\label{eq:sigmaseries} \sigma(\rho)=
\frac{v^2_R}{GR}
\frac{1}{2\pi}\sum\limits_k\hat{\sigma}_k\,J_0\br{
\omega_k\frac{\rho}{R}},\quad \rho\ll{}R.
\end{equation}
Despite that close to $R$ equation (\ref{eq:sigmaseries})
 is not a trustworthy approximation (the sum may become negative), it
almost perfectly agrees (usually for $\rho<0.6R$) with the surface
densities corresponding to several model rotation curves that were
cutting-off at different $R$.

\subsection{The iteration scheme}\label{sec:it}

From the above, it is evident that the total mass of the galactic
disk within the cutoff radius cannot be determined from its
rotation law only. However, if additional bounds for the surface
mass density are known outside the cutoff radius, one may find a
reliable global mass distribution by iterations.

As before we denote by $R$ the cutoff radius, to which a rotation
curve has been measured. We denote this part of rotation curve  by
$v_0$. The distribution of matter, which extends far beyond the
cutoff radius and which is  mostly composed of neutral and
molecular hydrogen, we denote by $\sigma_H$. If such data are not
available, one may try to devise a physically plausible profile of
baryonic matter outside $R$.

At the very beginning we define $u(x)\equiv{}v_0(R\,x)/v_0(R)$. We
calculate from (\ref{eq:sigmaseries}) an initial surface density
$\sigma_0$ where the corresponding spectral coefficients
$\hat{\sigma}^{(0)}_k$ are calculated from (\ref{eq:expanscoeff})
(we put $\sigma_0$ and $\hat{\sigma}^{(0)}_k$ in place of $\sigma$
and $\hat{\sigma}_k$, respectively). Let $R_1$ be the first value
of $\rho$ where $\sigma_0(\rho)=\sigma_H(\rho)$. The first
approximation $\sigma_1$ to the sought global surface density for
$\rho<R_1$
 is defined by setting $\sigma_1=\sigma_H$, and for $\rho>R_1$
we set $\sigma_1=\sigma_0$. With this $\sigma_1$ we begin the
first iteration step ($i=1$) which is described below for
arbitrary $i$.

\begin{description}
\item[The $i^{\mathrm{th}}$ iteration step] By setting $\sigma=\sigma_i$ in equation
(\ref{eq:vfromsig}) we calculate the corresponding approximation
$v=v_i$ to a global rotational velocity. Afterwards we find the
resulting discrepancy function $\delta{}v^2_i=v^2_0-v^2_i$ defined
for $\rho<R$. The function measures an approximation error in the
$i^{\mathrm{th}}$ iteration step and is used also to determine a
correction $\delta\sigma_i$ to the actual global surface mass
density $\sigma_{i}$. By substituting
$\delta{}v_i^2(x\,R)/\delta{}v_i^2(R)$ in place of $u^2(x)$ in
equation (\ref{eq:expanscoeff}) we find spectral coefficients
$\delta\hat{\sigma}^{(i)}_{k}$ where now $\hat{\sigma}_k$ stands
for  $\delta\hat{\sigma}_k^{(i)}$. Next we calculate for $\rho<R$
the correction $\delta\sigma_i$ from equation
$$\delta\sigma_i(\rho)=
\frac{\delta{}v_i^2(R)}{2\pi{}GR}
\sum\limits_k\delta\hat{\sigma}_k^{(i)}\,J_0\br{
\omega_k\frac{\rho}{R}},$$ which is analogous to equation
(\ref{eq:sigmaseries}). Let $R_{i+1}$ be the first value of $\rho$
where $\sigma_{i}(\rho)+\delta\sigma_i(\rho)=\sigma_H(\rho)$. The
corrected global surface mass density $\sigma_{i+1}$ is found by
setting $\sigma_{i+1}=\sigma_i+\delta\sigma_i$ for $\rho<R_{i+1}$
and $\sigma_{i+1}=\sigma_H$ for $\rho>R_{i+1}$.\end{description}
Now the $i+1$ iteration step follows. We repeat this procedure
until $\max(\delta{}v^2_{n})$ in the $n^{\mathrm{th}}$ iteration
becomes acceptably small.

\section{Application of the iteration method to the spiral galaxy NGC 4736}

As an example we choose the spiral galaxy NGC 4736. Its rotation
curve excludes the massive spherical halo, therefore the global
disk model should better approximate the gravitational field in
this galaxy than models assuming such a halo.

There is also the additional information about distribution of
hydrogen observed in the remote region of this galaxy, where the
rotation curve has not been determined. This provides the lower
bound for the surface density. No other luminous matter has been
observed there. The amount of hydrogen in this region is
comparably small. Its presence allows us to reconstruct the global
mass density and predict the rotation curve beyond the presently
measured range.

\subsection{Observational data} A high resolution rotation curve
of the galaxy was adopted from \citep{bib:rotcrv}. It was measured
out to $7'$ which corresponds to cutoff radius $R=10.38\kpc$ (the
distance of the galaxy is $D=5.1\mpc$ and inclination angle
$i=35^{\circ}$ as in \citep{bib:rotcrv}). The rotational velocity
has been measured with the maximum error $\delta{}v=\pm11\kms$
(attained at $\rho=5'$) and the average velocity is $168\kms$.
Thus the total dynamical mass should be determinable with the
accuracy better than $\approx10\%$. We use also the $50''$
resolution neutral hydrogen surface density published in
\citep{bib:hydrogen} (inclination was corrected from $40^{\circ}$
to $35^{\circ}$), which is consistent with \citep{bib:bosma}. A
$\beta$-band surface brightness $\lambda_{\beta}$, where
$\beta=I$, $V$ or $B$ is calculated from
$$\lambda_{\beta}\sq{\frac{L_{\beta,\sun}}{\mathrm{pc}^2}}=
\frac{\cos\br{i}}{\tan^2\br{\mathrm{arcsec}}}10^{0.4\br{
M_{\beta,\sun}-\mu_{\beta}-5}}.$$ It is done using the respective
luminosity profiles, corrected for galactic extinctions
$\mu_{\beta}\sq{\frac{\mathrm{mag}}{\mathrm{arcsec}^2}}$, taken
from \citep{bib:luminosity}. The reference absolute magnitudes for
the Sun are: $M_{I,\sun}=4.19$ \citep{bib:mi}, $M_{V,\sun}=4.83$
\citep{bib:5} and $M_{B,\sun}=5.43$ \citep{bib:mb}.

\subsection{Determining the global surface density and the global rotation curve}
Below we shall apply the iteration scheme described in section
\ref{sec:it}. We use the same notational conventions as introduced
there.

\begin{figure}[h!!!!]\centering
\includegraphics[width=\columnwidth]{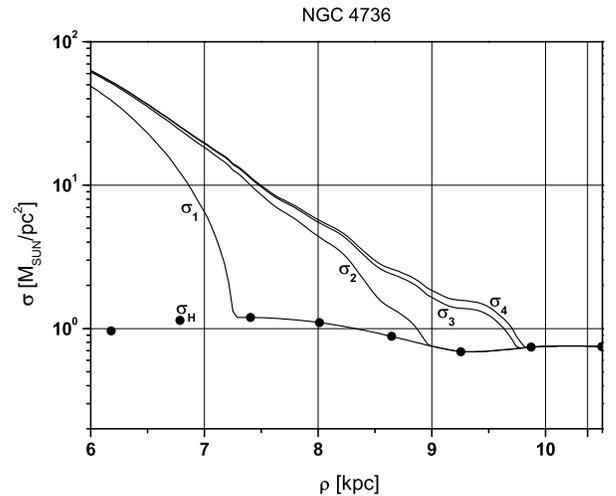}
\caption{\label{fig:f2} A fragment of surface mass density of
galaxy NGC 4736 obtained in consecutive iterations (solid lines
$\sigma_1$, $\sigma_2$, $\sigma_3$, $\sigma_4$) and the observed
column mass density of \hi{} ($\sigma_H$ -- solid circles).}
\end{figure}

\begin{figure}[h!!!!]\centering
\includegraphics[width=\columnwidth]{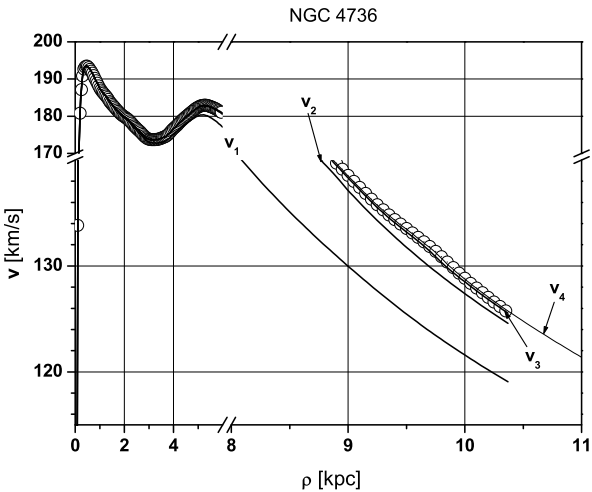}
\caption{\label{fig:f3}$v_1,v_2,v_3,v_4$ are rotational velocities
calculated from equation (\ref{eq:vfromsig}) for global surface
densities obtained in the four consecutive steps of iteration. The
thin line $v_4$ was shown also outside the observational range
(experimental points for NGC 4736 are indicated by circles).}
\end{figure}

\begin{figure}[h!!!!]\centering
\includegraphics[width=\columnwidth]{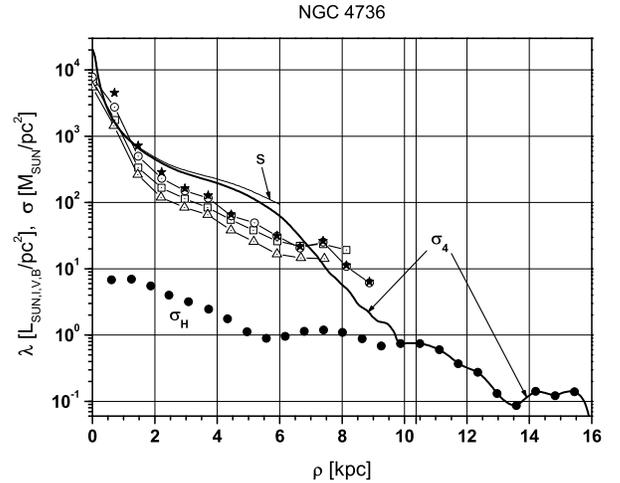}
\caption{Global surface mass density for NGC 4736 found in the
fourth iteration (thick solid line $\sigma_4$). This mass
distribution leads to rotation velocity that agrees perfectly with
the observed rotation curve. For comparison, the surface mass
density found in \citep{bib:5} is also shown (the thin line $s$).
The observed density profile of hydrogen \hi{} is marked with
solid circles. Luminosity profiles $\lambda(\rho)$ (at inclination
angle $35^{\circ}$) in the $B$-band (dotted triangles), the
$V$-band (dotted squares) and the $I$-band (dotted circles) were
corrected for galactic extinction \citep{bib:luminosity}. The
$I$-band luminosity was corrected also for the internal extinction
(marked with stars).\label{fig:f4}}
\end{figure}

\begin{figure}[h!!!!]\centering
\includegraphics[width=\columnwidth]{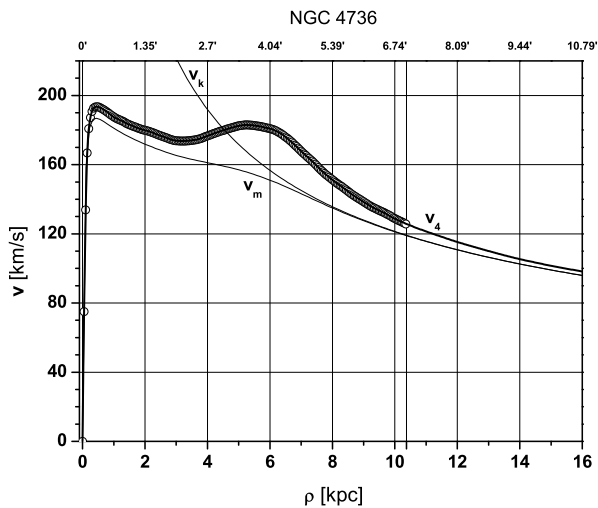}
\caption{\label{fig:f5}The observed rotation curve of galaxy NGC
4736 (circles) and the rotation curve (solid thick line $v_4$)
corresponding to surface mass density $\sigma_4$ found in the
fourth iteration and calculated from equation (\ref{eq:vfromsig}).
The solid thin line $v_k$ is the corresponding Keplerian asymptote
$\sqrt{GM_{TOT}/\rho}$ where
$M_{TOT}\approx3.43\times10^{10}M_{\sun}$ is the total galaxy
mass. The solid thin line $v_m$ is the rotational velocity which
one would obtain from a spherically symmetric mass distribution
with the same mass function as that for NGC 4736.}
\end{figure}

A third order interpolating function $u^2(x)=v_0^2(R\,x)/v_R^2$
($v_R\equiv{}v_0(R)$ $=125.6\kms$) is used to find the spectral
coefficients $\hat{\sigma}^{(0)}_k$. The $\mu_k$ coefficients in
(\ref{eq:expanscoeff}) fall off quickly to zero, eg.
$\mu_1/\mu_{125}\approx3\times10^{-6}$, and the functional
sequence of partial sums in (\ref{eq:sigmaseries}) converges
quickly. The summation can thus be terminated at some finite $k$
(we put $k=144$). This truncated sum is taken as the surface mass
density $\sigma_0(\rho)$. Next, we set
$\sigma_1(\rho)=\sigma_0(\rho)$ for $\rho<R_1$,
$\sigma_1(\rho)=\sigma_H(\rho)$ for $R_1<\rho<R_H$ and
$\sigma_1(\rho)=0$ elsewhere, where $\sigma_H(\rho)$ is the
azimuthally averaged neutral hydrogen surface density, $R_1$ was
defined in sec. \ref{sec:it} and $R_{H}=16.06\kpc>R$, cf. line
'$\sigma_1$' in figure \ref{fig:f2}.
%
%
%
%
The integrated mass  of $\sigma_1(\rho)$ is
$M_1\approx3.144\times10^{10}M_{\sun}$ and should not exceed the
total mass $M_{TOT}$ more than $0<M_{TOT}-M_1<\delta{}M_1$, where
$\delta{}M_1<2Rv_0(R)\mathrm{Max}|v_0-v_1|/G
\approx5\times10^9M_{\sun}\approx0.16M_1$. The corresponding first
approximation $v_1(\rho)$ to the global rotation curve agrees with
$v_0(\rho)$ very well for the small radii, where $\sigma_1(\rho)$
is reliable. However, there is still a discrepancy at greater
radii, cf. line $v_1$ in figure \ref{fig:f3}. In order to reduce
it, we find a correction $\delta\sigma_1(\rho)$ to
$\sigma_1(\rho)$ in the region $0<\rho<R$. The needed spectral
coefficients $\hat{\sigma}_k^{(1)}$ were determined by finding a
least square fit to a first order interpolation of
$\delta{}v_1^2(\rho)$, with $xJ_{1}(\omega_k\,x)$ used as basis
functions.\footnote{It is known from the theory of generalized
Fourier series, that in the theoretical limit of infinite number
of terms in the sum (\ref{eq:vseries}), the values of
$\hat{\sigma}_k$ determined by the least square method would be
equal to those calculated from equation (\ref{eq:expanscoeff})}
Having found $\delta\sigma_1(\rho)$ we calculate the corrected
surface density in the second approximation $\sigma_2(\rho)=
\sigma_1(\rho)+\delta\sigma_1(\rho)$, with the reservation we set
$\sigma_2(\rho)=\sigma_H(\rho)$ for $\rho>R_2$ and
$\sigma_2(\rho)=0$ for $\rho>R_H$ (the line '$\sigma_2$' in figure
\ref{fig:f2}). Now, the corresponding global rotation curve
$v_2(\rho)$ agrees with $v_0(\rho)$ almost perfectly (the line
'$v_2$' in figure \ref{fig:f3}),
but  it still needs to be corrected a little in the vicinity of,
$R$, since $\mathrm{Max}\br{1-v_2/v_0}\approx0.02$. The
corresponding mass is $M_2=3.385\times10^{10}M_{\sun}$ and is
underestimated, but $M_{TOT}-M_2<\delta{}M_2$, where
$\delta{}M_2<2R\,v_0(R)\,\mathrm{Max}|v_0-v_2|/G
\approx1\times10^9M_{\sun}\approx0.03M_2$. Similarly, we carry out
the third step, to find $\sigma_3(\rho)$ and $v_3(\rho)$. The
integrated mass in this step is $M_3=3.419\times10^{10}M_{\sun}$
and $\delta{}M_3\,\approx\,3.3\times10^8M_{\sun}\approx0.01M_3$.
The error has not changed substantially compared with
$\delta{}M_2$, therefore we terminate iterations after the fourth
step. In this way we obtain the global surface density
$\sigma_4(\rho)$ shown in figure \ref{fig:f4}.
%
%
It gives the total mass $M_{TOT}=3.43\times10^{10}M_{\sun}$ (it
includes the observed hydrogen \hi{} out to $R_H$), and the
corresponding rotation curve $v_4(\rho)$ -- its predicted
continuation beyond the cutoff radius can be tested with future
observations (see figure \ref{fig:f5}).
%
%
%
\subsection{Discussion}
In the previous section we have obtained the surface density in
galaxy NGC 4736 that is consistent with distribution of \hi{}
beyond the cutoff radius $R$, and perfectly agrees with the
observed rotation curve. Since \hi{} mass outside $R$ is only
$0.003M_{TOT}$, the results (e.g density profile) would not be
affected significantly if there was an amount of baryonic matter
outside $R$ comparable to that of \hi{}. The global surface mass
density of galaxy NGC 4736 found in the fourth iteration is shown
in figure \ref{fig:f4} (the '$\sigma_4$' line). The corresponding
global rotation curve calculated from equation (\ref{eq:vfromsig})
agrees perfectly with the experimental rotation curve $v_0$, see
figure \ref{fig:f5}).

\begin{figure}[h!!!!]\centering
\includegraphics[width=\columnwidth]{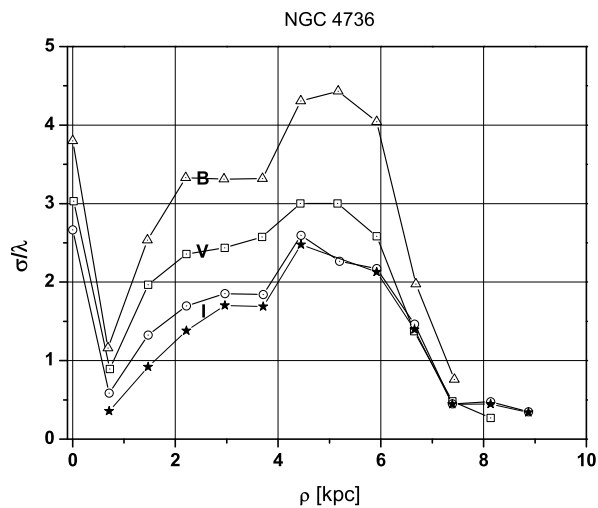}
\caption{\label{fig:f6} The predicted ratios of mass--luminosity
surface densities
 for galaxy NGC 4736 with luminosities corrected for galactic extinction in $B$-band
(triangles), $V$-band (squares), $I$-band (circles) and
 corrected additionally for internal extinction  $I$-band (marked with stars). }
\end{figure}

The $V$ and $B$-band luminosities are smaller than the $I$ band
luminosity $L_I$, cf. table \ref{tab:1} ($L_V\approx0.7L_I$ and
$L_B\approx0.5L_I$),
%
%
\begin{table}[h!!!!]
\begin{tabular}{|c|c|}
\hline
$M_{TOT}$& $3.43\times10^{10}M_{\sun}$ \\
$M_{HI}$& $6.00\times10^{8}M_{\sun}\Rightarrow 0.02M_{TOT}$\\
$L_{I} $& $2.87\times10^{10}L_{I,\sun}\Rightarrow M/L_I=1.2$\\
$L_{V} $& $2.14\times10^{10}L_{V,\sun}\Rightarrow M/L_V=1.6$\\
$L_{B} $& $1.54\times10^{10}L_{B,\sun}\Rightarrow M/L_B=2.2$\\
\hline
\end{tabular}
\caption{\label{tab:1} Physical parameters  obtained with the help
of the Iterative Spectral Method for spiral galaxy NGC 4736. (at
the distance $5.1[\mathrm{Mpc}]$ and inclination angle
$35^{\circ}$).}
\end{table}
%
which may indicate the presence of dust. The mass-to-light ratio
profile (surface density to surface luminosity ratio) is shown in
figure \ref{fig:f6}.
%
%
From the three bands, the $I$-band should be regarded as the most
reliable tracer of the luminous matter, as it most accurately maps
the amount of the radiated energy.  Luminosity in $V$- and
$B$-bands is reduced by the presence of dust, it also does not
include the radiation from low mass stars. For galaxy NGC 4736 the
$I$-band mass-to-light ratio is small: $M/L_I\approx1.2$. Also its
distribution, i.e. density to luminosity ratio which is of the
order of $1$ -- $2$ (cf. figure \ref{fig:f6}) is typical to stars
of slightly sub-solar mass. The total galaxy mass is surprisingly
small. It is comparable to the mass predicted in the framework of
the MOND model, which gives $3.21\pm0.09\times10^{10}M_{\sun}$, or
in the framework of the metric skew tensor gravity, which gives
$3.15\pm0.08\times10^{10}M_{\sun}$ \citep{bib:6}. We obtain
$M_{TOT}=3.43\times10^{10}M_{\sun}$ for the same rotation curve in
the framework of Newtonian physics.

\begin{figure}[h!!!!]\centering
\includegraphics[width=\columnwidth]{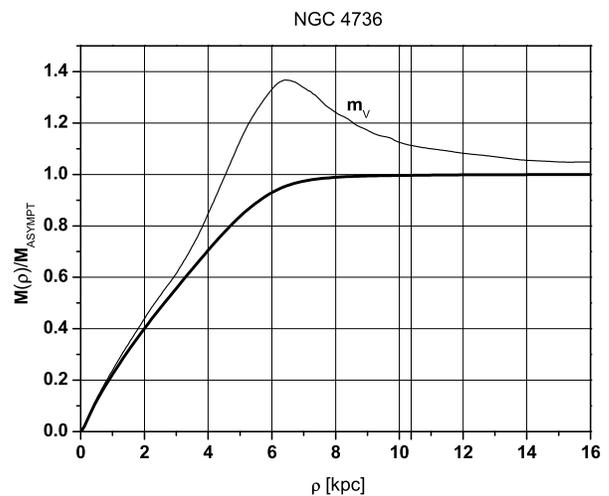}
\caption{\label{fig:f7} Mass function for NGC 4736 normalized with
respect to total mass $M_{TOT}=3.43\times10^{10}M_{\sun}$ (thick
solid line) compared with Keplerian mass function $m(\rho)$ ($m_V$
-- the thin solid line) defined as $\rho{}v_0^2(\rho)/(GM_{TOT})$,
where $v_0(\rho)$ is the observed rotation curve.}
\end{figure}

\noindent In figure \ref{fig:f7} is shown  the mass function
$$M_4(\rho)/M_{TOT}, \qquad  M_4(\rho)=2\pi\int_{0}^{\rho}\sigma_4(\xi)\xi\ud{\xi}.$$
It is shown together with the Keplerian mass function
$v_0^2(\rho)\rho/(GM_{TOT})$, corresponding to the observed
rotation curve $v_0(\rho)$.
%
%
%
%
The 'odd' behaviour of the latter function was our argument
against applying dark halo dominated models to galaxy NGC 4736.
Such a model used in \citep{bib:kent} predicted
$5.0\times10^{10}M_{\sun}$ for the total galaxy mass at a distance
$5.1\mpc$. The best fitting curve in that model gives rotational
velocity at the cutoff radius $1.2$ times greater than the
observed value. As this leads to overestimation of the total mass
by a factor of some $(1.2)^2$, one should expect rather
$5.0\times10^{10}M_{\sun}/(1.2)^2\approx3.47\times10^{10}M_{\sun}$
for the total mass, which is very close to our result obtained in
the global disk approximation. We note also a related and trivial
observation: Keplerian rotation law $\sqrt{M_4(\rho)G/\rho}$ for a
spherically symmetric matter distribution with the mass function
numerically equal to $M_4(\rho)$, gives smaller rotation
velocities at larger radii than the disk model. Also more matter
would be required by a spherical model to account for
observations, see figure \ref{fig:f5}.

\section{Conclusions}

The comparison of mass functions and rotation laws at the end of
the previous section, illustrates the fact that the models with
flattened mass distributions are more efficient than the commonly
used models assuming spherical halo. The former are better in
accounting both for high rotational velocities as well as for low
scale structure of rotation curves and with noticeably less amount
of matter than the latter (the relation between rotation and mass
distribution in the disk model is very sensitive for gradients of
a rotation curve). The use of the disk model is justified for
galaxies with rotation curves violating the sphericity condition.
This is necessary (although not sufficient) condition  for a
spherical mass distribution.\footnote{A thin disk with surface
density $\sigma(x)=(1+x^2)^{-3/2}/\br{2\pi}$ rotates with velocity
$v(x)=x(1+x^2)^{-3/4}$. This rotation low satisfies the sphericity
condition: $(xv^2(x))'=3x^2(1+x^2)^{-5/2}>0$.}

Rotation of the spiral galaxy NGC 4736 can be fully understood in
the framework of Newtonian physics. We have found a mass
distribution in the galaxy that agrees perfectly with its
high-resolution rotation curve, agrees with the $I$-band
luminosity distribution giving low mass-to-light ratio of $1.2$ in
this band at total mass of $3.43\times10^{10}M_{\sun}$,  and is
consistent with the amount of \hi{} observed in the remote parts
of the galaxy, leaving not much room (if any) for dark matter.
Remarkably, we have achieved this consistency without invoking the
hypothesis of a massive dark halo nor using modified gravities.

There exist a class of spiral galaxies, similar to NGC 4736, that
are not dominated by spherical mass distribution at larger radii.
Most importantly, in this region rotation curves should be
reconstructed accurately in order not to overestimate the mass
distribution. For a given rotation curve it can be easily
determined whether or not a spherical halo may be allowed at large
radii by examining the Keplerian mass function corresponding to
the rotation curve (the so called sphericity test).

By using complementary information of mass distribution,
independent of rotation curve, we overcame the cutoff problem for
the disk model, that for a given rotation curve, a mass
distribution could not be found uniquely as it was dependent on
the arbitrary extrapolation of the rotation curve.

\acknowledgements
One of the authors (MK) was  supported by the
Polish Ministry of Science and Higher Education, grant 1 P03D 005
28.

\newpage
\end{document}